\documentclass{egpubl}
\usepackage{pg2020w}
 
\WsWiP           

\usepackage[T1]{fontenc}
\usepackage{dfadobe}  

\usepackage{cite}  
\BibtexOrBiblatex
\electronicVersion
\PrintedOrElectronic
\ifpdf \usepackage[pdftex]{graphicx} \pdfcompresslevel=9
\else \usepackage[dvips]{graphicx} \fi

\usepackage{egweblnk} 

\usepackage{amsmath}

\title[RTSDF]{RTSDF: Generating Signed Distance Fields in Real Time for Soft Shadow Rendering}

\author[Tan et al.]
{\parbox{\textwidth}{\centering 
Tan Yu Wei\orcid{0000-0002-7972-2828}, 
Nicholas Chua, 
Clarence Koh, 
Anand Bhojan\orcid{0000-0001-8105-1739}
}
\\
{\parbox{\textwidth}{\centering School of Computing, National University of Singapore}}}

\begin{document}

\teaser{
 \includegraphics[width=\linewidth]{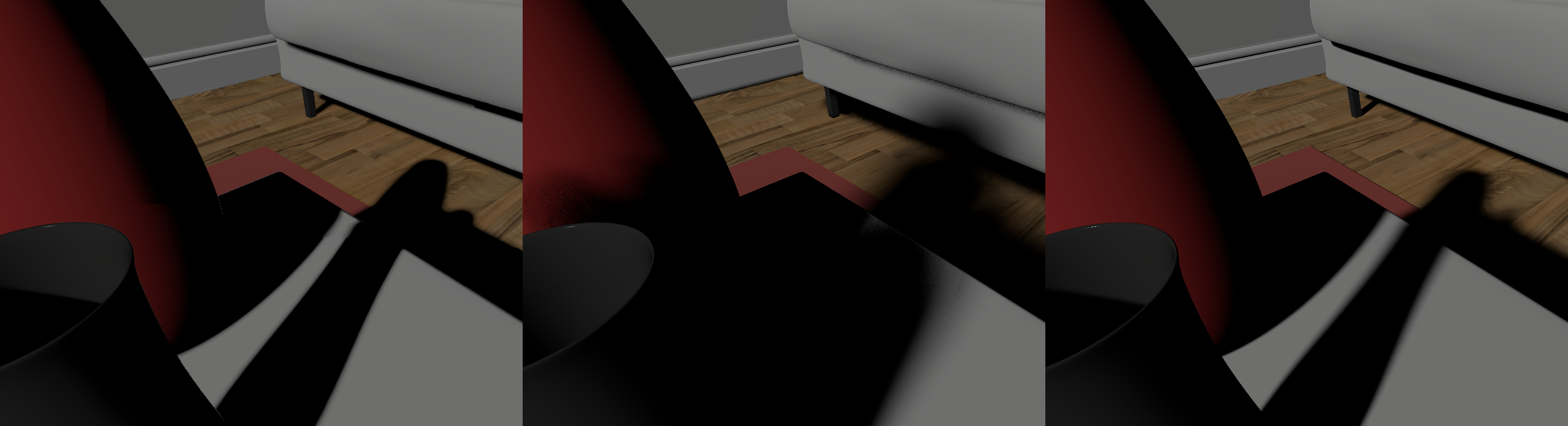}
 \centering
  \caption{From left to right: CSM (EVSM) (113 fps), jump flooded and ray-traced SDF (97 fps) and ground truth offline distributed ray tracing for soft shadows with one directional light source on \href{https://www.blendswap.com/blend/13491}{\textsc{The Modern Living Room}} (\href{https://creativecommons.org/licenses/by/3.0/}{CC BY}) with a GeForce RTX 2080 Ti}
\label{fig:teaser}
}

\maketitle
\begin{abstract}

Signed Distance Fields (SDFs) for surface representation are commonly generated offline and subsequently loaded into interactive applications like games. Since they are not updated every frame, they only provide a rigid surface representation. While there are methods to generate them quickly on GPU, the efficiency of these approaches is limited at high resolutions. This paper showcases a novel technique that combines jump flooding and ray tracing to generate approximate SDFs in real-time for soft shadow approximation, achieving prominent shadow penumbras while maintaining interactive frame rates.


\end{abstract}

\maketitle
 
\section{Introduction}

A brute force method to generate Signed Distance Fields (SDFs) offline is to trace rays in all directions and query the nearest surface \cite{Wright:2015:DOS}, which produces accurate surface representation with smooth and refined surfaces at high computational cost. SDFs can alternatively be approximated in real-time with jump flooding \cite{Rong:2006:JFG}, offering a voxelized scene representation that causes reconstructed surfaces to appear blocky. Hence, we propose a technique that combines the precision of ray tracing and the speed of jump flooding, generating more accurate SDFs for soft shadows of promising quality with real-time interactive performance.
\section{Design}

We first perform jump flooding to give the \emph{coarse} SDF or a fast approximation of the SDF of the scene. Next, we detect regions in the \emph{coarse} SDF near surfaces and apply brute force ray tracing with temporal accumulation in these areas to generate a \emph{fine} SDF for more accurate scene representation. Lastly, we use the \emph{fine} SDF to render soft shadows.

\subsection{Jump Flooding}

The jump flooding algorithm obtains information on the nearest seed to any point in space. Extending from 2D to 3D, we voxelize our scene and perform jump flooding with the voxels as seeds. 

At our chosen resolution for the \emph{coarse} SDF, we perform scene voxelization using the rasterization pipeline by rendering the scene with an orthographic camera, so that the screen coordinates can be easily transformed into voxel coordinates for updating the \emph{coarse} SDF texture. We perform this rendering from the positive x, y and z directions to voxelize the maximum area of projection of every primitive, minimizing holes in our voxelized mesh. The rasterization pipeline generates fragments for each primitive visible from the camera. To ensure that fragments for all primitives are generated, writes to the depth buffer are disabled to prevent depth culling. 

Jump flooding is then performed in parallel on the GPU to accelerate the calculations. The distance to the nearest voxel is stored in the \emph{coarse} SDF. As for the sign, we subtract a small experimentally-derived bias $\beta$ from the distance field, causing some surface points to be negative and effectively thickening the surfaces.

\subsection{Ray Tracing}

We first initialize a higher resolution \emph{fine} SDF. Adopting selective rendering like in adaptive frameless rendering, texels in the \emph{fine} SDF with distance to the nearest surface smaller than a user-defined variable \emph{d} in the \emph{coarse} SDF are ray-traced, generating more accurate distance values for the \emph{fine} SDF. \emph{d} can be used to tune the trade-off between performance and accuracy in SDF generation. When \emph{d} is large, more \emph{fine} SDF texels would be ray-traced as more \emph{coarse} SDF texels would be within distance \emph{d} from a surface point.

For each texel to be ray-traced, we then shoot rays in random directions from its world position to query its distance to the nearest surface. Taking this position as the origin, we generate random points on the surface of a sphere as our ray directions. At frame $t$, the distance traced by each ray to its closest hit is obtained and the smallest such distance for each texel is accumulated as $r_{t}$.

We use temporal accumulation over multiple frames to increase our effective sample count and converge to the final distance field. However, ghosting artifacts appear on the surface of moving objects as their previous position is not invalidated as the nearest surface. Hence, we apply a decay factor $\alpha$ to the distance field of the previous frame. Taking $f_{t - 1}$ as the \emph{fine} SDF at frame $t - 1$ and $c_{t}$ as the coarse SDF at frame $t$, the \emph{fine} SDF at frame $t$ is computed as:

\begin{equation}
  f_{t} = 
  \begin{cases} 
    \min(\alpha \cdot f_{t - 1} + (1 - \alpha) \cdot c_{t}, r_{t}), & \text{for } c_{t} \leq d \\
    c_{t}, & \text{for } c_{t} > d
  \end{cases} 
\end{equation}

We also temporally accumulate the number of front and back face hits. By convention, we take a negative sign to represent a point enclosed by the nearest surface. Hence, the \emph{fine} SDF is set to be negative if more rays hit back as compared to front faces.

\subsection{Soft Shadows}

We then raymarch the \emph{fine} SDF to approximate the visibility term and generate soft shadows \cite{Wright:2015:DOS}. However, we need to account for no intersection and prevent indefinite iteration. Hence, we terminate the raymarching when a maximum number of steps has been taken or the distance to the nearest surface is smaller than a value $\epsilon$. $\epsilon$ is minimally one voxel size to prevent self-occlusion. 

When shadows are parallel to the light direction, the raymarching uses a relatively constant step size that produces patterned shadows. Hence, we also jitter the ray origin towards the light and apply temporal anti-aliasing to stabilize noise introduced by the jitter.

We linearly sample and not point sample the SDF to achieve smoother and less blocky surface representation. However, due to the low resolution of the SDF, banding artifacts appear within shadow penumbras as alternating regions of high and low occlusion. Hence, we perform triangulation between current and previous raymarch sample points \cite{Aaltonen:2018:GCS}. Restricting the maximum step size helped to smooth out the occlusion amount between neighbouring pixels and remove the rest of the banding, but increased the number of texture samples required to calculate occlusion.
\section{Discussion}

We evaluate our method against commonly used CSM \cite{Engel:2006:CSM} with EVSM filtering \cite{Lauritzen:RAS:2008} for real-time soft shadows. As seen in \autoref{fig:teaser}, we attain more prominent penumbras and smoother shadow boundaries. However, our penumbras are larger than that of the ground truth due to the \emph{coarse} SDF $\beta$ and soft penumbra widening shadows technique \cite{Aaltonen:2018:GCS} which we plan to improve. 

Due to low resolution, jump flooding voxels corresponding to thin surfaces produces gaps in the \emph{coarse} SDF. Although our ray tracing step helps to fill in these gaps for the \emph{fine} SDF, the large raymarch $\epsilon$ of minimally one voxel size also causes thin surfaces to be stepped through entirely, resulting in holes appearing in shadow umbras. Hence, we hope to apply shadow mapping in these regions instead \cite{Wright:2015:DOS} for quality and performance enhancements.

\section*{Acknowledgements}

This work is supported by the Singapore Ministry of Education Academic Research grant T1 251RES1812, “Dynamic Hybrid Real-time Rendering with Hardware Accelerated Ray-tracing and Rasterization for Interactive Applications”. 

\bibliographystyle{eg-alpha-doi}
\bibliography{rtsdf}

\end{document}